\documentclass[aps,twocolumn,showpacs]{revtex4}
\usepackage{graphicx}
\def\bsigma{\mbox{\boldmath $\sigma$}}

\begin{document}

\title{Interplay of intra- and interband absorption in a disordered graphene}
\author{F. T. Vasko}
\email{ftvasko@yahoo.com}
\author{V. V. Mitin}
\affiliation{Department of Electrical Engineering, University at Buffalo, Buffalo, NY 1460-1920, USA}
\author{V. Ryzhii}
\author{T. Otsuji}
\affiliation{Research Institute of Electrical Communication, Tohoku University, Sendai 980-8577 and Japan Science and Technology Agency, CREST, Tokyo 107-0075, Japan }
\date{\today}

\begin{abstract}
The absorption of heavily doped graphene in the terahertz (THz) and mid-infrared (MIR) spectral regions is considered taking into account both the elastic scattering due to finite-range disorder and the variations of concentration due to long-range disorder. Interplay between intra- and  interband transitions is analyzed for the  high-frequency regime of response, near the Pauli blocking threshold. The  gate voltage and temperature dependencies of the absorption efficiency are calculated. It is demonstrated that for typical parameters, the smearing of the interband absorption edge is determined by a unscreened part of long-range disorder while the intraband absorption is determined  by finite-range scattering. The latter yields the spectral dependencies which deviate from those following from  the Drude formula. The obtained  dependencies are in good agreement with recent experimental results. The comparison of the results of our calculations with the experimental data provides a possibility to extract the  disorder characteristics. 
\end{abstract}
\pacs{72.80.Vp, 73.50.Bk, 78.67.Wj}
\maketitle

\section{Introduction}
Both gapless (massles) energy band structure and peculiarities of scattering mechanisms determine the response of graphene on the in-plane polarized radiation in the THz and MIR spectral regions. Such a response is caused by both  the direct intersubband transitions and the intraband transitions of free carriers accompanied by scattering processes (the Drude mechanism), see the reviews \cite{1,2,3} and references therein. Contributions of these two mechanisms in heavily-doped graphene are described by the inter- and intraband dynamic conductivities,  $\sigma_{\rm inter}$ and $\sigma_{\rm intra}$, respectively, given by 
\begin{equation}
{\rm Re}\sigma_{\rm inter}\approx\frac{{e^2 }}{4\hbar}\theta (\hbar\omega -2\varepsilon_F ), ~~~ \sigma_{\rm intra}(\omega )\approx\frac{\sigma_F}{1+i\omega\tau_F }  ,
\end{equation}
where $\varepsilon_F$ and $\tau_F$ stand for the Fermi energy and the momentum relaxation time, $\sigma_F =(e^2 /4\hbar )\varepsilon_F \tau_F /\hbar$ is the static conductivity, and $\theta(\varepsilon)$ is a unity step-like fuction. Here we have neglected a weak imaginary contribution ${\rm Im} \sigma_{\rm inter}$ \cite{4} and the smearing of the interband absorption edge associated with the finiteness of the temperature, $T$,  and the collisional damping of the electron spectrum. When such a damping can be neglected  at $T = 0$, the  expression for ${\rm Re} \sigma_{\rm inter}$ describes an abrupt jump at photon energy $\hbar\omega =2\varepsilon_F$ due to the Pauli blocking effect. The interband and intraband contributions are of the same order of magnitude  at $\omega\tau_F\sim\sqrt {4\varepsilon_F\tau_F /\pi\hbar -1}$.  Therefore, the  overlap of these contributions disappears if $\varepsilon_F\tau_F /\hbar\geq 20$ (see Fig. 1a). Thus, the  description of interplay between both contributions is necessary in the THz (or MIR, depending on the doping level) spectral region for the typical samples with $\hbar /\varepsilon_F\tau_F\geq 0.2$. As shown below, the quasi-classical description of the intraband absorption and, therefore, the expression for  $\sigma_{\rm intra} (\omega )$ based on the Drude formula is not valid if $\hbar\omega$ exceeds temperature of carriers \cite{5,6}. This is  because of the variation of the relaxation time as a function of the electron energy  over energy intervals $\sim\hbar\omega$, where the intraband transitions take place (see Fig. 1b).
\begin{figure}[tbp]
\begin{center}
\includegraphics[scale=0.6]{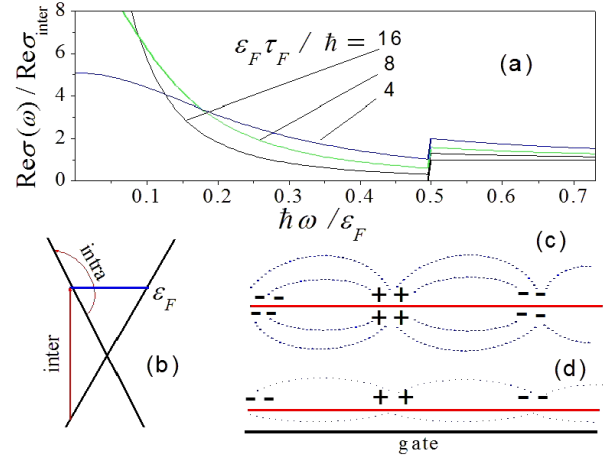}
\end{center}
\addvspace{-0.5 cm}
\caption{(a) Spectral dependencies of ${\rm Re}\sigma (\omega )$  described by simplified Eqs. (1) for different doping levels determined by $\varepsilon_F \tau_F /\hbar$. (b) Intra- and interband transitions between linear branches of gapless energy spectra. (c) Screening of non-uniform charge in ungated graphene with field distribution shown by dashed curves. (d) Gated graphene structure with unscreened charge and distribution of weak field shown by dotted curves.}
\end{figure}

The constant value for the interband conductivity ${\rm Re}\sigma_{\rm inter}$ associated with  the frequency-independent absorption was observed for undoped suspended graphene in the visible range of the spectrum.  \cite{7} . By analogy with electrical transport in field-effect transistors, MIR spectroscopy  allows for the control of the Pauli blocking effect using the electrical gating~\cite{8}.  In  sufficiently heavily  doped graphene, the Drude mechanism of absorption was observed near the interband threshold, up to the MIR spectral region.~\cite{9,10,11}. Despite of a series of theoretical studies (see Refs. 12-14 and references therein), which have treated the problem of intra- and interband absorption at different levels of approximation, there are two drawbacks in the interpretation of the above-listed experimental data. First, the quantum description of the intraband absorption was not connected with the scattering parameters determined from the conductivity measurements and, second, an inhomogeneous smearing of the interband threshold due to long-range disorder was not considered. Thus, a re-examination of the intra- and interband absorption processes in a disorded graphene is timely now.

In this paper, we present a general consideration which is based on the Kubo formula and takes into account both the scattering due to finite-range disorder and {\it a long-range unscreened variation of concentration}. The latter contribution appears to be essential because the  screening of the potential spatial variations associated with the disorder with the in-plane scales longer than (or comparable to) the distance to gate is suppressed. This mechanism is schematically illustrated in Figs. 1c and 1d.  As a result, the inhomogeneous mechanism of smearing of the interband absorption at $\hbar\omega\sim 2\varepsilon_F$ should be taken into account, together with the thermal smearing of the Fermi distribution and broadening of the spectral density due to the elastic scattering on the finite-range disorder. The dynamic conductivity in the high-frequency range (if $\omega\tau_F >1$) is expressed via  the Green's function, which is averaged over the finite-range disorder, and the carrier distribution averaged over the long-range non-screened disorder. The averaged characteristics of Green's function were obtained using the phenomenological description of the static conductivity similarly to Ref. 15.  Considering the interplay of the intra- and interband contributions, we analyze the spectral, temperature, and gate voltage dependencies of the  absorption efficiency. The obtained results are compared with the recent experimental data.~\cite{10,11}. The fitting procedure demonstrates that the disorder characterization, including the evaluation of the parameters of high-energy scattering and long-range inhomogeneties, can be performed using the THz and MIR spectroscopy.
 
The paper is organized as follows. In the next section (Sec. II), we present the basic equations which describe the inter- and intraband mechanisms of absorption and consider unscreened long-range disorder in gated graphene. In Sec. III we analyze the spectra of relative absorption and their dependencies on the temperature disorder and as well as on the doping conditions. The last section includes the discussion of experimental data and the approximations used, and the conclusions. 

\section{Basic Equations}
We start with consideration of the dynamic conductivity based on the Kubo formula which takes into account the elastic scattering and the unscreened long-range disorder. Since the  absorption is determined by the real part of conductivity, ${\rm Re}\sigma_{\omega}$, we neglect the ${\rm Im}\sigma_{\omega}$ contribution. \cite{4}

\subsection{Dynamic conductivity}
Absorption of the THz (MIR ) radiation propagating along the normal to the graphene layer is described by the real part of dynamic conductivity, 
\cite{6} given by
\begin{eqnarray}
{\rm Re}\sigma_{\omega}=\frac{2\pi e^2}{\omega L^2}\int {dE} \left( f_E  - f_{E + \hbar \omega }\right) ~~~ \\
\times\left\langle\sum\limits_{\alpha\alpha '}|(\alpha |\hat{\upsilon}_\| |\alpha ' )|^2\delta (E-\varepsilon_\alpha )\delta (E -\varepsilon_{\alpha '}+\hbar\omega) \right\rangle \nonumber .
\end{eqnarray}
Here $L^2$ is the normalization area, $f(\varepsilon_\alpha )$ is the equilibrium distribution over the states $|\alpha )$ with energies $\varepsilon_\alpha$  and $|(\alpha |\hat{\upsilon}_\| |\alpha ' )|^2$ is the matrix element of the in-plane velocity operator $\hat{\upsilon}_\|$ between the states $\alpha$ and $\alpha '$ connected by the energy conservation law described by the $\delta$-function. The $\alpha$-states are defined by the eigenvalue problem $(\hat h+V_{\bf x})\Psi_{\bf x}^{(\alpha )}=\varepsilon_\alpha\Psi_{\bf x}^{(\alpha )}$ expressed via the single-particle Hamiltonian $\hat h$, a random potential $V_{\bf x}$, and the two-component wave function $\Psi_{l{\bf x}}^{(\alpha )}$ with $l=$1,2. The statistical averaging of Eq. (2) over the random potential is denoted as $\langle\ldots\rangle$. In order to perform this averaging, we introduce the spectral density function 
\begin{equation}
A_\varepsilon  \left( {l{\bf x},l'{\bf x}'} \right) = \sum\limits_\alpha  {\delta \left(\varepsilon -\varepsilon_\alpha \right)\Psi_{l{\bf x}}^{(
\alpha )} \Psi _{l'{\bf x}'}^{(\alpha )*} }  .
\end{equation}
Thus, ${\rm Re}\sigma_\omega$ given by Eq. (1) is transformed into
\begin{eqnarray}
{\rm Re}\sigma_\omega =\frac{2\pi (e\upsilon )^2}{\omega L^2}\int {dE} \left( f_E -f_{E + \hbar \omega }\right) \int {d{\bf x}_1 } \int {d{\bf x}_2 } ~~  \\
\times \left\langle {\rm tr}\left[ {\hat A}_E({\bf x}_1 ,{\bf x}_2 )\hat{\sigma}_\| \hat A_{E+\hbar\omega}({\bf x}_2 ,{\bf x}_1 )\hat{\sigma}_\|
\right] \right\rangle  , \nonumber
\end{eqnarray}
where ${\rm tr}\ldots$ means trace over the isospin variable, $\upsilon =10^8$ cm/s is the characteristic velocity, and $\hat{\sigma}_\| =(\hat{\bsigma}\cdot{\bf e})$ is written through  the $2\times 2$ Pauli matrix $\hat{\bsigma}$, and the polarization ort ${\bf e}$.

Further, the spectral density function is expressed through the exact retarded and advanced Green's functions, $\hat{\cal G}_E^R \left( {{\bf x}_1 ,{\bf x}_2 } \right)$ and $\hat{\cal G}_E^A \left( {{\bf x}_1 ,{\bf x}_2 } \right)$, as follows $\hat A_E\left({\bf x}_1 ,{\bf x}_2\right)=i\left[ {\hat{\cal G}_E^R \left( {{\bf x}_1 ,{\bf x}_2 } \right) - \hat{\cal G}_E^A \left( {{\bf x}_1 ,{\bf x}_2 } \right)} \right]/2\pi$. These Green's functions are governed by the standard equation
\begin{eqnarray}
\left[ E-i0-\upsilon (\hat{\bsigma}\cdot {\bf p}_1)-u_{{\bf x}_1}-v_{{\bf x}_1}\right] \nonumber \\
\times{\cal G}_E^R \left( {{\bf x}_1 ,{\bf x}_2 } \right) =\hat 1\delta \left( {{\bf x}_1  - {\bf x}_2 } \right)  ~~~~~
\end{eqnarray}
which is written through the potential energy with separated contributions from finite- and long-range disorder, $u_{{\bf x}_1}$ and $v_{{\bf x}_1}$ labeled as $fr$- and $lr$- respectively. The averaging $\left\langle\cdots \right\rangle$ in Eq. (4) should also be separated as $\left\langle\cdots \right\rangle_{lr}$ and $\left\langle\cdots\right\rangle_{fr}$. It is convenient to introduce the variables $({\bf x}_1+{\bf x}_2 )/2={\bf x}$ and ${\bf x}_1-{\bf x}_2  = \Delta {\bf x}$ in Eqs. (4) and (5). Using these variables and neglecting a weak contribution of $\nabla v_{\bf x}$ we obtain $v_{{\bf x}+\Delta {\bf x}/2}\simeq v_{\bf x}$. As a result, after the replacement $E\rightarrow E+v$ one can separate the averaging over $fr$- and $lr$-disorder contributions and Eq. (4) takes the form: 
\begin{eqnarray}
{\rm Re}\sigma_\omega  =\frac{2\pi}{\omega}(e\upsilon )^2\int {dE} \left( \left\langle f_{E+v}  \right\rangle_{lr}- \left\langle 
f_{E+v+\hbar\omega }\right\rangle_{lr} \right)   ~~ \\
\times \int d\Delta {\bf x} \left\langle {\rm tr}\left[ \hat A_E({\bf x}_1 ,{\bf x}_2 )\hat{\sigma}_\| \hat A_{E+\hbar\omega}({\bf x}_2 ,{\bf x}_1 )\hat{\sigma}_\|
\right] \right\rangle_{fr}  , \nonumber
\end{eqnarray}
where we take into account that $\left\langle\ldots\right\rangle_{fr}$ depends only on $|\Delta{\bf x}|$  and use $\int {d{\bf x}}  = L^2$.
 
We consider below the high-frequency spectral region, $\omega\tau_F\gg 1$, when the correlation function $\left\langle\ldots\right\rangle_{fr}$ can be factorized through the averaged spectral functions $\hat A_{\Delta {\bf x}E} \equiv \left\langle {\hat A_E ({\bf x}_1 ,{\bf x}_2 )} \right\rangle _{fr}$. The averaging over $fr$-disorder gives the contribution:
\begin{equation}
\int {d\Delta {\bf x}} \left\langle {\rm tr}\left[  \cdots  \right]\right\rangle _{sr}  = L^{ - 2} \sum\limits_{\bf p} {{\rm tr}\left( {\hat A_{pE} \hat \sigma _{||} \hat A_{pE + \hbar \omega } \hat \sigma _{||} } \right)} .
\end{equation}
Here the spectral densities $\hat A_{pE}$ should be written through the averaged Green's functions, $\hat G_{pE}^A=\hat G_{pE}^{R+}$ and $\hat G_{pE}^R$, which are given by the matrix expression:
\begin{eqnarray}
\hat G_{pE}^R  = \left[ {E - \upsilon (\hat \sigma  \cdot {\bf p}_1 ) - \hat \Sigma _{pE}^R } \right]^{-1}  , \\
\hat\Sigma_{pE}^R =\zeta_{pE}^R  + \frac{{(\hat{\bsigma}\cdot {\bf p})}}{p}\eta _{pE}^R ,  \nonumber
\end{eqnarray}
where the self-energy-function is determined in the Born approximation through the functions
\begin{eqnarray}
\left| {\begin{array}{*{20}c}
   {\zeta_{pE}^R }  \\
   {\eta _{pE}^R }  \\
\end{array}} \right| = \int {\frac{{d{\bf p}_{\bf 1} }}{{(2\pi \hbar )^2 }}} W_{|{\bf p} - {\bf p}_1 |} \\
\times \left| {\begin{array}{*{20}c}
   {\frac{1}{{E + i0 - \upsilon p_1 }} + \frac{1}{{E + i0 + \upsilon p_1 }}}  \\
   {\frac{{({\bf p} \cdot {\bf p}_1 )}}{{pp_1 }}\left( {\frac{1}{{E + i0 - \upsilon p_1 }} - \frac{1}{{E + i0 + \upsilon p_1 }}} \right)}  \\
\end{array}} \right|  .   \nonumber 
\end{eqnarray}
Using the Fourier transformation of $\left\langle u_{{\bf x}_1}u_{{\bf x}_2} \right\rangle_{fr}=\overline{u}^2\exp [-({\bf x}_1  -{\bf x}_2 )^2/2l_{fr}]$, one obtains in (9) the Gaussian correlation function $W_{\Delta p}$ expressed through the strength of potential and correlation length, $\overline{u}$ and $l_{fr}$. Because $\hat G_{pE}^R$ depends only on the matrix $(\hat{\bsigma}\cdot{\bf p})$, the averaged spectral density of Eq. (7) takes the form
\begin{eqnarray}
\hat A_{{\bf p}E}  = \frac{{A_{pE}^{( + )}  + A_{pE}^{( - )} }}{2} + \frac{{(\hat \sigma  \cdot {\bf p})}}{{2p}}\left( {A_{pE}^{( + )}  - A_{pE}^{( - )} } \right) \\
A_{pE}^{( \pm )}  = \frac{i}{{2\pi }}\left( {\frac{1}{{\varepsilon _{pE}  \mp \upsilon _{pE} p}} - \frac{1}{{\varepsilon_{pE}\pm\upsilon_{pE} p}}}\right)  , \nonumber
\end{eqnarray}
where we introduced the renormalized dispersion law and velocity, $\varepsilon_{pE}=E-\zeta _{pE}^R$ and $\upsilon _{pE}  = \upsilon \left( {1 + \eta _{pE}^R /\upsilon p} \right)$, respectively.

For the averaging over $lr$-disorder in Eq. (6) we introduce the population factor $\left\langle f_{E+v}\right\rangle_{lr}=\int\limits_{-\infty }^\infty  {dE'} \phi_{E-E'}f_{E'}$ with the kernel $\phi_{E-E'}$ determined by the characteristics of disorder, see Sect. IIB. Thus, based on the two assumptions employed (smoothness of $lr$-potential, $\nabla v\to 0$, and high-frequency approach, $\omega\tau_F\gg 1$) we obtain ${\rm Re}\sigma_\omega$ written through the triple integral:
\begin{eqnarray}
{\rm Re}\sigma_\omega   = \frac{{\pi (e\upsilon )^2 }}{{\omega L^2 }}\int {dE\int {dE} '} \phi _{E - E'} \left( {f_{E'}  - f_{E' + \hbar \omega } } \right) \nonumber  \\
\times\sum\limits_{\bf p} {\left( {A_{pE}^{( + )}  + A_{pE}^{( - )} } \right)} \left( {A_{pE + \hbar \omega }^{( + )}  + A_{pE + \hbar \omega }^{( - )} } \right)  , ~~~~~
\end{eqnarray}
where the contributions $\propto A_{pE}^{(\pm )}A_{pE}^{(\pm )}$ and $\propto A_{pE}^{(\pm )}A_{pE}^{(\mp )}$ correspond to the intra- and interband absorption processes, respectively. Similar expressions for the homogeneous case, without $lr$-contribution, were considered in Refs. 13.

\subsection{Non-screened long-range disorder }
Here we evaluate the square-averaged potential of $lr$-disorder which is caused by a built-in random potential $w_{\bf x}$. We take into account the screening effect in the gated structure with the graphene sheet placed at $z=0$ on the substrate of the width $d$ and the static dielectric permittivity $\epsilon$, see Fig. 1d. The electrostatic potential $V_{{\bf x}z}$ is defined by the Poisson equation with the induced charge at $z=0$ determined by the concentration of electrons in the heavily-doped graphene: $n_{\bf x}=(4/L^4)\sum_{\bf p}\theta (\varepsilon_F -\varepsilon_{{\bf x}p})=[(\varepsilon_F -v_{\bf x})/\sqrt{\pi}\upsilon\hbar ]^2$. The dispersion law $\varepsilon_{p{\bf x}}=\upsilon p +v_{\bf x}$ is written here through the screened potential $v_{\bf x}=w_{\bf x}+V_{{\bf x}z=0}$. Performing the Fourier transformation over $\bf x$-plane one obtains the second order equations for $V_{{\bf q}z}^<$ and $V_{{\bf q}z}^>$ which correspond to the substrate, $0>z>-d$, and to the upper half-space, $z>0$:
\begin{equation}
\left( {\frac{{d^2 }}{{dz^2 }} - q^2 } \right)\left| {\begin{array}{*{20}c}
 {V_{{\bf q}z}^ >}  \\  {V_{{\bf q}z}^ <  }  \\  \end{array}} \right| = 0, 
	~~~\begin{array}{*{20}c}    {z > 0}  \\  {0 > z >  - d}  \\ \end{array}  .	
\end{equation}
The jump of electric field at $z=0$ is defined by the charge distribution over graphene ($z\to 0$)
\begin{eqnarray}
\left. \frac{dV_{{\bf q}z}^>}{dz} \right|_{z = 0}-\epsilon \left. \frac{dV_{{\bf q}z}^<}{dz} \right|_{z = 0}  =  - \Psi _{\bf q} , \nonumber \\
\Psi_{\bf q}=4\pi e^2 \int\limits_{L^2 } d{\bf x} e^{-i{\bf qx}} n_{\bf x} ,
\end{eqnarray}
so that the potential remains homogeneous at $z=0$: $V_{{\bf q}z=0}^<  =V_{{\bf q}z = 0}^>\equiv V_{{\bf q}z=0}$. The boundary condition at $z=-d$ is defined by the in-plane homogeneous back-gate voltage $V_g$ as follows: $V_{{\bf q}z=-d}^< =V_g\delta_{{\bf q},0}$. The last boundary condition is the requirement that the potential $V_{{\bf q}z}^ >$ vanishes at $z\to\infty$.  

The solution of this electrostatic problem can be written through $\exp (\pm qz)$ and $\Psi_{\bf q}$ which depends on $V_{{\bf q}z=0}$ through the concentration $n_{\bf x}$. Solving the system of the boundary conditions, we find that the screening potential at $z=0$ should satisfy the requirement:
\begin{eqnarray}
V_{{\bf q}z=0}= -\Psi _{\bf q}dK(qd),  \\
K(y)=\frac{\sinh y}{y(\sinh y +\epsilon\cosh y)} , ~~\nonumber
\end{eqnarray}
where the dimensionless function $K$ describes the gate-induced quenching of $V_{{\bf q}z=0}$ due to the fixed potential at $z=-d$. According to Eq. (13), $\Psi_{\bf q}$ is the nonlinear function of the screened potential written through $n_{\bf x}$ and the electrostatic problem given by Eqs. (12) and (13) is transformed into the integral equation for $V_{{\bf q}z=0}$ determined by (14). Further, we consider the case of heavily-doped graphene, when $\varepsilon_F\gg |v_{\bf x}|$ and the electron-hole puddles are absent, so that $n_{\bf x}\approx n_F (1-2v_{\bf x}/\varepsilon_F )$ is written through the averaged concentration $n_F$. As a result, the linearized dependency between $\Psi_{\bf q}$ and the screened potential $v_{\bf q}$ takes place
\begin{equation}
\Psi _{\bf q}  \approx 4\pi e^2 n_F\left( {\delta _{{\bf q},0}  - \frac{{2v_{\bf q} }}{\varepsilon_F}} \right)  .
\end{equation}
 Therefore, Eqs. (14) and (15) give the linear relation between $V_{{\bf q}z=0}=v_{\bf q}-w_{\bf q}$ and the Fourier component of the screened potential $v_{\bf q}$. The solution for $v_{\bf q}$ takes form:
\begin{equation}
v_{\bf q} =\frac{w_{\bf q}}{1+ 8(e/\hbar\upsilon )^2\varepsilon_F dK(qd)}, 
\end{equation}
where screening is suppressed (i.e. $v_{\bf q}\to w_{\bf q}$) if $d\to 0$ and the complete screening ($|v_{\bf q}|\ll |w_{\bf q}|$) takes place at $d\to\infty$.

The averaging over $lr$-disorder in the distribution $\left\langle f_{E+v}\right\rangle_{lr}$ is performed after the Fourier transform of $f_E$ with the use of the relation $\left\langle \exp \left[ iv_{\bf x}\tau /\hbar  \right] \right\rangle_{lr} =\exp\left[ -(\overline{v}\tau /\hbar )^2 /2 \right]$, where $\overline{v}=\sqrt{\left\langle v_{\bf x}^2 \right\rangle_{lr}}$ is the averaged $lr$-potential. Under the inverse Fourier transformation (i.e. the integration over time) of the averaged distribution one obtains the result $\left\langle f_{E+v}\right\rangle_{lr}=\int dE'\phi_{E-E'}f_{E'}$ which is written through the Gaussian kernel
\begin{equation}
\phi_{\Delta E}=\exp\left[ -(\Delta E/\overline{v})^2 /2\right] /(\sqrt{2\pi}\overline{v})   
\end{equation}
with the half-width $\overline{v}$.
Using Eq. (16) one can write $\overline{v}$ through the averaged built-in potential $w_{\bf q}$. We consider here a simple case of the Gaussian distribution of $w_{\bf x}$ described by the correlation function
\begin{equation}
\left\langle w_{\bf q} w_{{\bf q}'} \right\rangle_{lr}  = L^2 \delta_{{\bf q} + {\bf q}',0} 2\pi l_{lr}^2 \overline{w}^2 \exp \left[ -(ql_{lr} )^2 /2\right] ,
\end{equation}
which is written through the correlation length $l_{lr}$ and the averaged built-in potential $\overline{w}=\sqrt{\left\langle w_{\bf x}^2 \right\rangle_{lr}}$. Using Eqs. (18) and (19) one obtains
\begin{equation} 
\frac{\overline{v}}{\overline{w}}=\left\{\int\limits_0^\infty  {\frac{{dxxe^{-x^2 /2} }}{{\left[ 1+g_F (d/l_{lr}) K(xd/l_{lr} )\right]^2 }}}\right\}^{1/2} ,
\end{equation}
and this ratio depends on $d/l_{lr}$ and on the coupling constant, $g_F =(8e^2/\hbar\upsilon )l_{lr}/\lambda_F \propto\varepsilon_F l_{lr}$, written through the Fermi wavelength $\lambda_F =\upsilon\hbar /\varepsilon_F$.  If $d/l_{lr}\ll 1$ one obtains $\overline{v}/\overline{w}\to 1$ and in the region $d/l_{lr}\geq 1$ one obtains $\overline{v}/\overline{w}\simeq\sqrt{(1+\epsilon )/g_F}$ which is independent on $d/l_{lr}$. In Fig. 2 we plot the function (19) for the case of SiO$_2$ substrate if the coupling constant $g_F\simeq$10 - 100 corresponds to the concentration range $\sim 4\times 10^{11}$ - $4\times 10^{13}$ cm$^{-2}$ and if $l_{lr}\sim$50 nm. With increasing of $n_F$ or $l_{lr}$ at $g_F >100$, the relation $\overline{v}/\overline{w}\simeq\sqrt{(1+\epsilon )/g_F}$ appears to be valid for the interval of $d/l_{lr}>0.1$.
\begin{figure}[tbp]
\begin{center}
\includegraphics[scale=0.6]{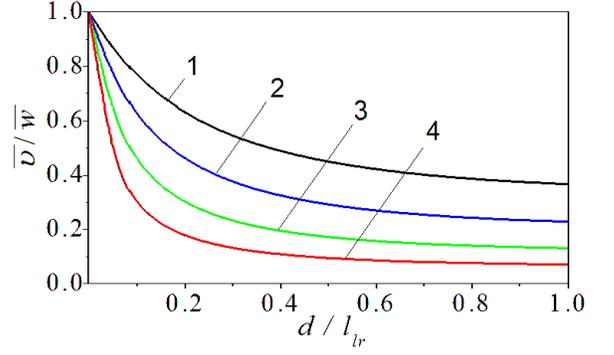}
\end{center}
\addvspace{-0.5 cm}
\caption{Dimensionless averaged potential $\overline{v}$ versus $d/l_{lr}$ for different $g_F =$12.5 (1), 25 (2), 50 (3), and 100 (4). }
\end{figure}

\section{Relative absorption}
In this section we consider the absorption of radiation propagated along the normal to graphene sheet placed on semi-infinite substrate with the high-frequency dielectric permittivity $\kappa$. Here we analyze the relative absorption coefficient 
\begin{equation}
\xi_\omega\simeq\frac{{16\pi{\rm Re}\sigma_\omega  }}{{\sqrt \kappa\left( {1+\sqrt\kappa}\right)^2 c}}
\end{equation}
neglecting the second-order corrections with respect to the parameter $4\pi |\sigma_\omega |/c\sqrt \kappa$. Note that the reflection and transmission coefficients, $R_\omega$ and $T_\omega$ (these values are connected by the energy conservation requrement: $R_\omega +T_\omega +\xi_\omega =1$), involve the first-order correction $\propto 4\pi |\sigma_\omega |/c\sqrt\kappa$ and the ${\rm Re}\sigma_\omega$ contributions should be involved for description of $R_\omega$ and $T_\omega$.

\subsection{Spectral dependencies}
First, we consider $\xi_\omega$ given by Eq. (20) after substitution the dynamic conductivity  determined by Eqs. (9) - (11). Performing the multiple integrations in ${\rm Re}\sigma_\omega$, one obtains the spectral, gate voltage, and temperature dependencies of relative absorption. For calculations of the renormalized energy spectra and velocity defined by Eq. (9) we use here the typical parameters of $fr$-disorder corresponding  to the sample with the maximal sheet resistance $\sim$3.5 k$\Omega$ and the correlation length $l_{fr}\sim$7.5 nm, see \cite{15} for details. Level of long-range disorder $\overline{v}$, correspondent to $l_{lr}\gg\lambda_F$, is determined by the ratio $d/l_{lr}$, see Fig. 2; here we consider heavily-doped graphene because a more complicated analysis is necessary for the case when electron-hole puddles are formed. \cite{16} 
\begin{figure}[tbp]
\begin{center}
\includegraphics[scale=0.55]{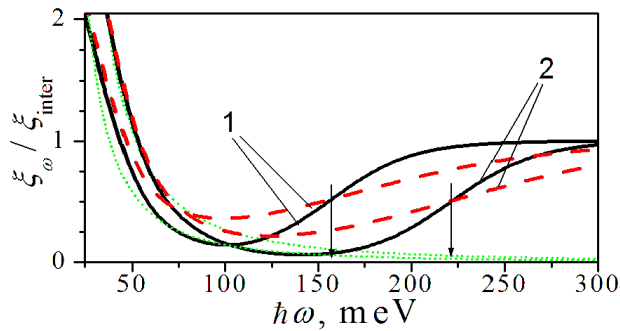}
\end{center}
\addvspace{-0.5 cm}
\caption{Spectral dependencies of relative absorption in homogeneous graphene (normalized to $\xi_{\rm inter}=\xi_{\omega\to\infty}$) at temperatures 77 K and 300 K (solid and dashed curves, respectively) for concentrations $5\times 10^{11}$ cm$^{-2}$ (1) and  $10^{12}$ cm$^{-2}$ (2). Arrows correspond to the doubled Fermi energies and dotted curves are $\propto\omega^{-2}$ asymptotics for Drude absorption. }
\end{figure}

The spectral dependencies of $\xi_\omega$ normalized to the high-frequency interband contribution, $\xi_{\rm inter}$, are plotted in Fig. 3 for the homogeneous graphene case, $\overline{v}=0$, at concentrations $5\times 10^{11}$ cm$^{-2}$ and  $10^{12}$ cm$^{-2}$ for temperatures $T=$77 K and 300 K. Similarly to the schematic spectra in Fig. 1a, one can separate contributions from intra- and interband transitions at $\hbar\omega\sim$100 meV and $\sim$150 meV for lower- and higher-doped samples with the minimal value of $\xi_\omega /\xi_{\rm inter}\sim$0.15 and $\sim$0.08, respectively. A visible deviation from the Drude spectral dependence ($\propto\omega^{-2}$) takes place due to quantum character of intraband transitions if $\hbar\omega\geq T$. Different signs of these deviations, which depend on $\hbar\omega$, $T$, and $n_F$, appear due to complicated energy and momentum dependencies in Eqs. (9) and (10). The smearing of interband absorption around the Fermi energies ($2\varepsilon_F$ are marked by arrows) is defined by both the scattering-induced broadening of the spectral density (10) and thermal effect, c.f. solid and dashed curves in Fig. 3. 
\begin{figure}
\begin{center}
\includegraphics[scale=0.55]{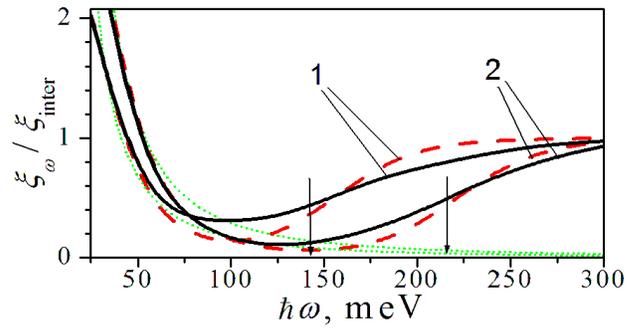}
\end{center}
\addvspace{-0.5 cm}
\caption{ The same as in Fig. 3 at $T=$77 K taking into account inhomogeneous smearing of threshold due to non-screened variations of concentrations around $5\times 10^{11}$ cm$^{-2}$ (1) and  $10^{12}$ cm$^{-2}$ (2). Dashed curves are plotted for the homogeneous case, $\overline{v}=$0. }
\end{figure}

In addition, the inhomogeneous smearing of interband absorption threshold due to non-screened long-range variations of concentration is essential, as it is shown in Fig. 4 for $T=$77 K. Here we considered a strong $lr$-disorder case with $\overline{v}\simeq$30 meV for $n_F =5\times 10^{11}$ cm$^{-2}$ and $\overline{v}\simeq$20 meV for $n_F =10^{12}$ cm$^{-2}$ (we used the dependencies of Fig. 2 with $\overline{w}\simeq$50 meV and $d/l_{lr}\leq$0.1). Similarly to the thermally-induced effect, the unscreened disorder effect results both in smearing of the interband absorption threshold and in enhancement of the minimal absorption in the region between intra- and interband contributions.

\subsection{Comparison with experiment }
We turn now to comparison of our calculations with the recent experimental data, \cite{10,11} where both the spectral dependencies of relative absorption in large-area samples and the static conductivity $\sigma$ versus gate voltage $V_g$ were measured. Using the phenomenological momentum relaxation rate $\nu_p$ suggested in Ref. 15 and the experimental data for hole conductivity from Refs. 10 and 11, one can fit the dependency $\sigma (V_g )$ as it is shown in Fig. 5. This approach gives us the $fr$-disorder scattering parameters for energies $vp\leq$200 meV and we can extrapolate these data up to $\sim$400 meV energies, which are necessary for description of mid-IR response measured in Refs. 10 and 11.
\begin{figure}
\begin{center}
\includegraphics[scale=0.6]{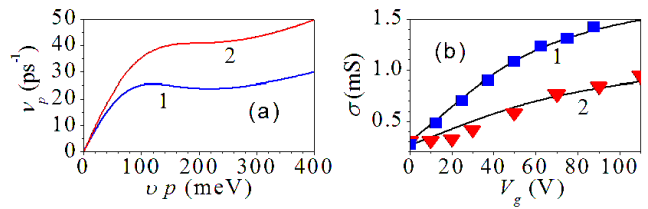}
\end{center}
\addvspace{-0.5 cm}
\caption{Fitting of experimental data from Refs. 10 and 11 marked as 1 and 2: (a) momentum relaxation rate $\nu_p$ versus energy $\upsilon p$ and (b) sheet conductivity versus gate voltage. Squares and triangles are experimental points from Refs. 10 (1) and 11 (2), respectively. }
\end{figure}

In Fig. 6 we plot the spectral dependencies of relative absorption measured in Ref. 10 for graphene with hole concentrations $\sim$2.2, $\sim$4.3, and $\sim$5.8$\times 10^{12}$ cm$^{-2}$ at $T=$100 K. Solid curves are plotted for the case of homogeneous sample with the scattering parameters determined from the transport data shown in Fig. 5. These dependencies are in agreement with experiment both in the intraband absorption region (at $\hbar\omega\leq$150 meV) and above threshold of interband absorption, at $\hbar\omega >2\varepsilon_F$. But the absorption in the intermediate region $\hbar\omega\sim$200 - 300 meV appears to be suppressed in comparison to the  experimental data. The minimal absorption increases if we take into account the unscreened $lr$-disorder contribution with $\overline{v}\sim$80 meV, 75 meV, and 60 meV for curves 1, 2, and 3, respectively. Such values of $\overline{v}$ are realized if $\overline{w}$ is comparable to $\varepsilon_F$ and $d/l_{lr}\leq 0.1$, i.e. micrometer-scale of inhomogeneities takes place, see Fig. 2.  Thus, we have obtained a reasonable agreement with the experimental data. But the smearing of the interband threshold is stronger in comparison to experimental data because the Gaussian model does not describe a real $lr$-distribution. A more accurate description is possible with the use of $lr$-disorder parameters taken from additional structure measurements.
\begin{figure}[tbp]
\begin{center}
\includegraphics[scale=0.55]{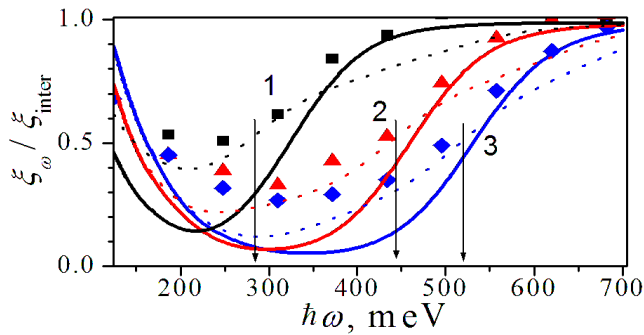}
\end{center}
\addvspace{-0.5 cm}
\caption{Fitting of relative absorption spectra (normalized to $\xi_{\rm inter}$) to conditions of Ref. 10 at gate voltages -30 V (1), -60 V (2), and -80 V (3). Solid and dotted curves correspond to homogeneous and inhomogeneous (with $\overline{w}\sim$150 meV) samples. Squares, triangles, and diamonds are experimental points for the cases 1, 2, and 3, respectively. Arrows correspond to the doubled Fermi energies. }
\end{figure}

Similar fitting of $\xi_\omega /\xi_{\rm inter}$ measured in Ref. 11 for hole concentrations $\sim$4.3, $\sim$6.5, and $\sim$8.6, and 11$\times 10^{12}$ cm$^{-2}$ at $T=$300 K is presented in Fig. 7. We plot the spectral dependencies both for the homogeneous sample (solid curves) and for the inhomogeneous sample taking into account the strong $lr$-disorder (dotted curves) with $\overline{v}\sim$120 meV, 105 meV, 90 meV, and 75 meV for curves 1, 2, 3, and 4, respectively. Now, good agreement with experimental data for interband absorption takes place (see the above discussion of Fig. 6 about the $lr$-disorder contribution) while the intraband absorption appears to be 2 - 3 times stronger in comparison to the disorder-induced contributions calculated. An additional contribution may appear due to relaxation via emission of optical phonons of energy $\hbar\omega_0$. Such a channel of intraband absorption is allowed if  $\varepsilon_F >\hbar\omega_0$ and the rate of transitions is proportional to the density of states at $\varepsilon_F -\hbar\omega_0$. As a result, this process may be essential for the conditions of Ref. 11, where $\varepsilon_F\geq$250 meV, while this mechanism should be negligible at lower concentrations used in Ref. 10. More detailed experimental data, for the spectral range 200 - 400 meV at temperatures $\leq$300 K, are necessary in order to verify this contribution. 
\begin{figure}[tbp]
\begin{center}
\includegraphics[scale=0.55]{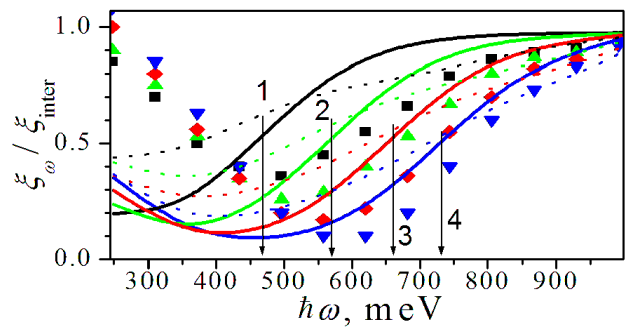}
\end{center}
\addvspace{-0.5 cm}
\caption{ The same as in Fig. 6 for conditions of Ref. 11 at gate voltages -30 V (1), -60 V (2), -90 V (3), and -120 V (4). Squares, triangles, diamonds, and inverse triangles are experimental points for the cases 1, 2, 3, and 4, respectively. }
\end{figure}

Overall, the comparison performed here demonstrates good agreement with experimental data, which opens a way for verification of scattering mechanisms and long-range disorder parameters. A complete verification can be performed under complex analysis of transport (conductivity versus $V_g$ and $T$) and spectroscopic data together with a structural (e. g., STM) measurements of long-range disorder.

\section{CONCLUSIONS}
We have examined the spectral, concentration (gate voltage), and temperature dependencies of the relative absorption due to intra- and interband transitions in graphene taking into account both elastic scattering and non-screened long-range variations of concentration. Smearing of the interband absorption edge, deviation from the Drude spectral dependence due to quantum regime of intraband transitions, and interplay of these two contributions are demonstrated. The results are in agreement with the recent experimental data obtained for  heavily-doped graphene. \cite{10, 11} This analysis allows to extract disorder parameters in the high-energy region (up to $\hbar\omega$ above the Fermi energy).

Let us discuss the assumptions used in the calculations performed. The main restriction of the results is due to consideration of the high-frequency spectral region, when ${\rm Re}\sigma_\omega$ can be written through the matrix product of the averaged spectral density function (10). More complicate description, based on the Bethe-Salpeter equation, is necessary for the low-frequency region, which is beyond of our consideration. We also restrict ourselves by the phenomenological models for the finite-range disorder scattering and for the long-range variations of concentration. It is enough for description of spectral, gate voltage, and temperature dependencies of $\xi_\omega$ because they are expressed through simple correlation functions which are similar for any microscopic nature of disorder. In addition, a long-range random strain \cite{17} may affect on the interband transitions in large-area samples. The	scattering processes caused by phonons and carrier-carrier collisions should be analyzed under calculations of the intra- and interband absorption in clean samples. A possible contribution of optical phonons is discussed in Sect. IIIB (more detailed analysis is beyond of the scope this paper) while the interaction with acoustic phonons is negligible. The comparison with experiment was performed here for the hole doping case and electron-hole asymmetry reported in Refs. 10 and 11 is unclear. According to Refs. 18, the Coulomb-induced renormalization of response and scattering processes give a weak contribution for typical disorder levels. This disagreement requires an additional investigation.

To conclude, we believe that our results open a way for characterization of scattering processes and long-range inhomogeneities using THz and MIR spectroscopy. More important, that effect of disorder on THz and MIR response is the main factor which determines characteristics of different devices (e.g., lasers \cite{19} and photodetectors \cite{20}) in this spectral region. We believe that our study will stimulate a further investigation of these device applications.

\section*{ACKNOWLEDGMENT}
This research is supported by the NSF-PIRE-TeraNano grant (USA), by the JSPS Grant in Aid for Specially Promoting Research (No. 23000008, Japan), and by the JSPS Core-to-Core Program (No. 23004, Japan).


\begin{thebibliography}{99}
\bibitem{4}
The imaginary contribution ${\rm Im}\sigma_{\omega}$ is expressed through ${\rm Re}\sigma_{\rm inter}$ with the use of the Kramers-Kronig relation. Using a cut-off frequency correspondent to the width of valence band, $\varepsilon_m$, one obtains a weak contribution: ${\rm Im} \sigma _\omega\sim  (\hbar\omega /2\pi\varepsilon_m )e^2/\hbar$.
\end{thebibliography}
\end{document}